\documentclass{iau} 
\usepackage{graphicx}

\def \ga{\mathrel{\mathchoice   {\vcenter{\offinterlineskip\halign{\hfil
$\displaystyle##$\hfil\cr>\cr\sim\cr}}}
{\vcenter{\offinterlineskip\halign{\hfil$\textstyle##$\hfil\cr
>\cr\sim\cr}}}
{\vcenter{\offinterlineskip\halign{\hfil$\scriptstyle##$\hfil\cr
>\cr\sim\cr}}}
{\vcenter{\offinterlineskip\halign{\hfil$\scriptscriptstyle##$\hfil\cr
>\cr\sim\cr}}}}}

\title[Extragalactic maser surveys] 
{Extragalactic maser surveys}

\author[C. Henkel, J.~E. Greene, \& F. Kamali ]  
{C. Henkel$^{1,2}$, J.-E. Greene$^3$ \and F. Kamali$^1$}

\affiliation{$^1$Max-Planck-Institut f{\"u}r Radioastronomie, Auf dem H{\"u}gel 60, 53121 Bonn, Germany \\ 
$^2$Astron. Dept., King Abdulaziz University, P.O. Box 80203, Jeddah 21589, Saudi Arabia \\ 
$^3$Department of Astrophysical Sciences, Princeton University, Princeton, NJ 08544, USA \\
email: {\tt chenkel@mpifr-bonn.mpg.de}}

\pubyear{2017}
\volume{336}  
\jname{Astrophysical Masers: Unlocking the Mysteries of the Universe}
\editors{A. Tarchi, M.J. Reid \& P. Castangia, eds.}
\begin{document}

\maketitle

\begin{abstract}
Since the IAU (maser-)Symposium 287 in Stellenbosch/South Africa (Jan. 2012), 
great progress has been achieved in studying extragalactic maser sources. Sensitivity has 
reached a level allowing for dedicated maser surveys of extragalactic objects. These 
included, during the last years, water vapor (H$_2$O), methanol (CH$_3$OH), and formaldehyde 
(H$_2$CO), while surveys related to hydroxyl (OH), cyanoacetylene (HC$_3$N) and ammonia (NH$_3$) 
may soon become (again) relevant. Overall, with the upgraded Very Large Array (VLA), the Atacama 
Large Millimeter/submillimeter Array (ALMA), FAST (Five hundred meter Aperture Synthesis Telescope)
and the low frequency arrays APERTIF (APERture Tile in Focus), ASKAP (Australian Square Kilometer 
Array Pathfinder) and MeerKAT (Meer Karoo Array Telescope), extragalactic maser studies are expected 
to flourish during the upcoming years.  The following article provides a brief sketch of past 
achievements, ongoing projects and future perspectives.  
\end{abstract}

\firstsection 

\section{The past few decades}

Maser lines allow us to detect and to localize tiny hotspots with exceptional activity across the Universe. 
These can be used to highlight regions with enhanced star formation, allow for determinations of proper 
motion out to distances of several Mpc, and help us to constrain the morphology and distance of galaxies. 
Before addressing recent or ongoing extragalactic maser surveys, here we briefly summarize achievements 
obtained till the time of the Stellenbosch meeting. We then proceed with the Local Group of galaxies (Sect.\,2),
H$_2$O megamaser projects (Sects.\,3--6), the search for a correlation between H$_2$O and OH megamasers
(Sect.\,7), formaldehyde (Sect.\,8), methanol, HC$_3$N, and HCN (Sect.\,9), and future perspectives (Sect.\,10).  

Extragalactic 1.7\,GHz (18\,cm) OH masers were detected as early as in the mid seventees (e.g., 
\cite[Gardner \& Whiteoak 1975)]{GardnerWhiteoak75}, soon followed by the detection of a bright 22\,GHz 
H$_2$O maser in M~33 by \cite[Churchwell et al. (1977)]{Churchwell77}. While line luminosities were high 
by Galactic standards, they were not surpassing their more local cousins by many orders of magnitude. 
This drastically changed, when the first ``megamasers'' were detected, this time in opposite order, 
starting with 22\,GHz H$_2$O in NGC~4945 (\cite[Dos Santos \& Lepine 1979)]{DosSantosLepine79} and 
followed by 1.7\,GHz OH in Arp~220 (\cite[Baan et al. 1982)]{Baan82}. During the following years the 
interest mainly focused on OH, because these megamasers were rapidly associated with a so far not well 
known class of galaxies, the ULIRGs (UltraLuminous InfraRed Galaxies). This culminated in the large 
survey carried out by \cite[Darling \& Giovanelli (2002)]{DarlingGiovanelli02}, leading to a total 
number of $\approx$100 detected OH megamaser sources. With respect to 22\,GHz H$_2$O megamasers, it 
took more than 15 years until it became clearer what they represented and in which class of galaxies 
they could be found. However, even today detection rates are typically below 10\%. Crucial discoveries
were the detection of satellite lines well off the systemic velocity in NGC~4258 (\cite[Nakai et al. 
1993)]{Nakai93}, the discovery of velocity drifts of the near systemic maser components of NGC~4258 
(\cite[Haschick et al. 1994]{Haschick94}; \cite[Greenhill et al. 1995]{Greenhill95}), the mapping 
of the H$_2$O maser features in NGC~4258 (\cite[Miyoshi et al. 1995)]{Miyoshi95}, and the discovery 
that 22\,GHz H$_2$O masers are most commonly found in Seyfert~2 and LINER (Low Ionization Nuclear 
Emission Line Region) galaxies (\cite[Braatz et al. 1996, 1997)]{Braatz96}. All this indicated 
that some of the luminous water vapor masers, the so-called disk-masers, are forming Keplerian 
parsec or even sub-parsec scale accretion disks around their galaxy's supermassive nuclear engines, 
allowing for a determination of the disk's geometry, the mass of the nuclear engine, and the angular 
diameter distance to the parent galaxy. 

Another molecular transition, the 4.8\,GHz $J$ = 1 K-doublet line of formaldehyde (H$_2$CO).  was also 
proposed to be masing, greatly surpassing any Galactic counterpart in luminosity, Commonly seen in absorption, 
quasi-thermal 4.8\,GHz emission lines form at densities of at least several 10$^5$\,cm$^{-3}$, while maser 
emission is rarely seen in the Galaxy (e.g., \cite[Ginsburg et al. 2015)]{Ginsburg15}. 4.8\,GHz H$_2$CO 
emission was detected toward Arp~220, IC~680, and IR\,15107+0724 (\cite[Baan et al. 1986]{Baan86}; \cite[Baan 
et al. 1993]{Baan93}; \cite[Araya et al. 2004)]{Araya04} and, in the initial paper, the line from Arp~220 
was interpreted as a maser likely amplifying the non-thermal redio background of the source.

\section{The Local Group}
The Large and Small Magellanic Clouds (LMC and SMC) allow for studies of masers commonly observed in 
the Galaxy, but under conditions of low metallicity and strong UV-radiation fields. During the past 
few years, \cite[Breen et al. (2013)]{Breen13}, \cite[Imai et al. (2013)]{Imai13}, and \cite[Johanson 
et al. (2014)]{Johanson14}, mainly using the Australia Compact Array (ATCA), were successfull in 
finding new maser spots related to star formation. In the LMC, now we have 27 H$_2$O masers originating 
from 15 regions of star formation, and 3 such masers from late type stars. With the study of \cite[Breen 
et al. (2013)]{Breen13}, four new 22\,GHz H$_2$O masers were detected in the SMC, thus tripling the 
number obtained three decades before by \cite[Scalise \& Braz (1982)]{Scalise82}. Most of these sources 
can greatly help, through proper motion measurements, to constrain the individual rotation of the two 
galaxies as well as to measure their orbital motion around the Milky Way.

Proper motion is also the main motivation to systematically measure the less conspicious members of the 
Local Group at 6.7 and 22\,GHz, searching for methanol and water vapor masers using the Sardinia Radio 
Telescope (SRT). Results from this ongoing project are summarized by the contribution of A. Tarchi. 

In the Andromeda galaxy (M~31) \cite[Sjouwerman et al. (2010)]{Sjouwerman10} and \cite[Darling 
(2011)]{Darling11} detected first 6.7\,GHz CH$_3$OH and 22\,GHz H$_2$O masers, respectively. After this 
encouraging start, however, no new maser sources could be identified during the following years (see 
\cite[Darling et al. 2016]{Darling16}, who surveyed $\approx$500 additional sources with compact 
24$\mu$m emission in M~31). The unknown proper motion of M~31 is the main obstacle on the way to a 
basic understanding of Local Group dynamics. Therefore, detecting many such maser lines with suitable 
flux density for interferometric studies is of high importance.

\begin{figure}[b]
\hspace{0.75cm}
\includegraphics[width=13.5cm]{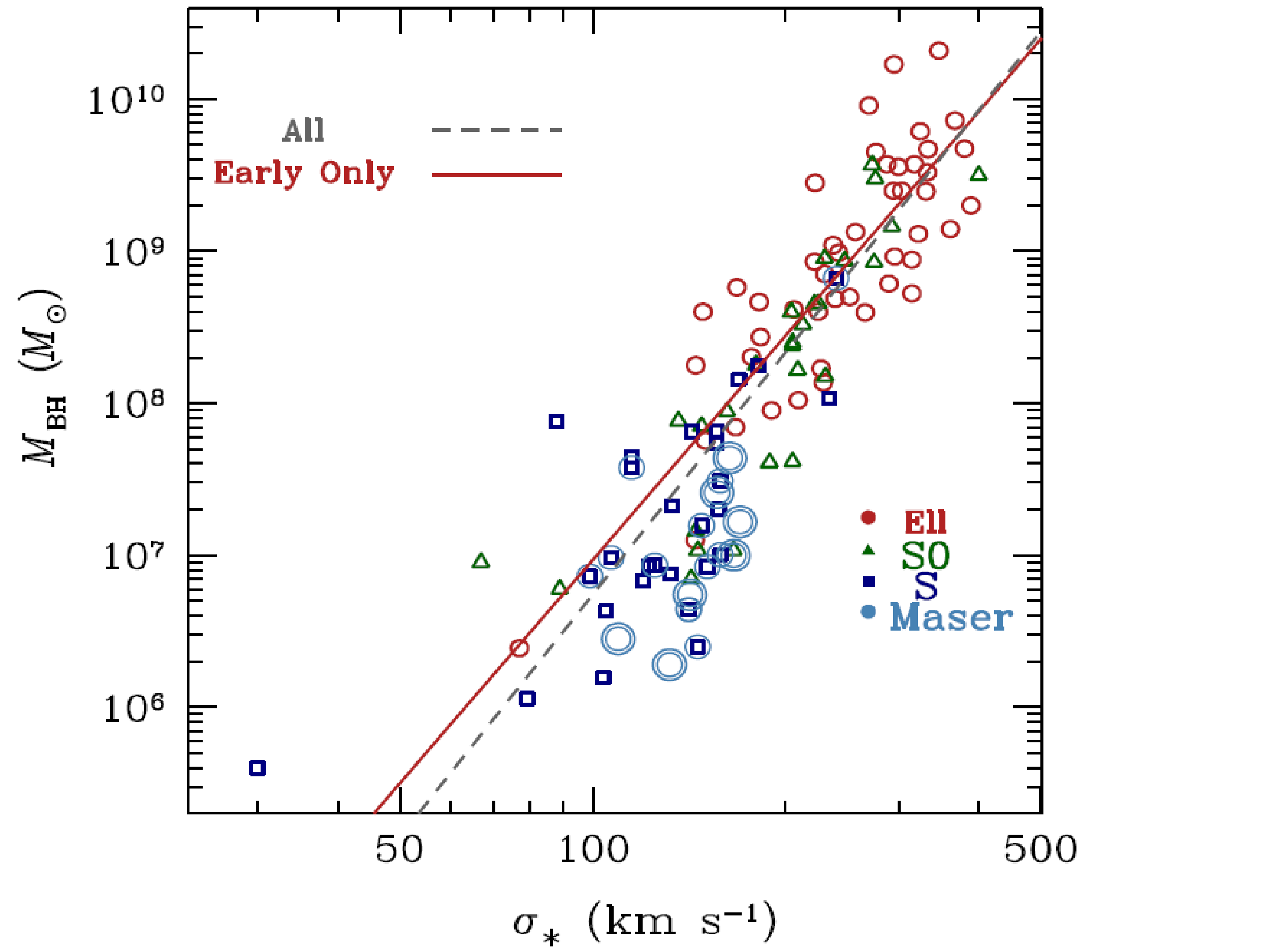}
\caption{Relationship between stellar velocity dispersion and central mass ($M_{\rm BH}$, taken from 
\cite[Greene et al.  2016]{Greene16}). Dashed line: Fit to the entire sample. Solid line: Early type 
galaxies only. Small open circles: ellipticals; small open triangles: lenticulars; small open squares: 
spirals; single circles surrounding an open square or double circles: disk-maser galaxies.}
\label{jenny}
\end{figure}

\section{H$_2$O megamaser detection surveys}
Already before the Stellenbosch meeting, dedicated 22\,GHz H$_2$O maser surveys have been carried out. 
These include unsuccessful searches for Fanaroff-Riley I (FR~I) galaxies (\cite[Henkel et al. 
1998)]{Henkel98} and relatively nearby low luminosity type~I and type~II QSOs (\cite[Bennert et al. 
2009]{Bennert09}; \cite[K{\"o}nig et al. 2012)]{Koenig12} as well as the detection of maser emission in an 
FR~II galaxy (\cite[Tarchi et al. 2003]{Tarchi03}). Additional surveys lead to the detection of a
type~2 QSO at redshift $z$ = 0.66 (\cite[Barvainis \& Antonucci 2005]{Barvainis05}) and to the 
detection of a gravitationally lensed type~I QSO at $z$ = 2.64 (\cite[Impellizzeri et al. 
2008]{Impellizzeri08}; \cite[Castangia et al. 2011]{Castangia11}; \cite[McKean et al. 2011)]{McKean11}.

Searching for 22\,GHz H$_2$O masers mainly toward Narrow-Line Seyfert~1 galaxies (NLS1s), which appear 
to contain relatively low mass nuclear emgines but luminosities compatible with their broad-line counterparts, 
\cite[Tarchi et al. (2011)]{Tarchi11} reported two new detections, thus leading to a total of five known
masers in this type of galaxies. The more recent survey by \cite[Hagiwara et al.  (2013a)]{Hagiwara13a} 
did not yield new positive results. 

Another survey lacking new detections, but also with high relevance for our understanding of the H$_2$O megamaser
phenomenon and the calibration of nuclear mass determinations, has been carried out by \cite[van den Bosch
et al. (2016)]{vandenBosch16}. They observed galaxies, where the gravitational sphere of influence of the 
central engine is extended enough to be resolvable by present day optical or near infrared (NIR) facilities. 
The detection of H$_2$O disk-masers in such galaxies would have a great impact. The radio data would provide 
a reliable black hole mass, which could then be compared with those derived by other potentially less accurate 
methods using optical or NIR data. The non-detections imply that NGC~4258 was and still remains the only such 
calibrator, where nuclear masses derived from H$_2$O and other methods, applicable to a larger number of galaxies,
could be compared. Furthermore, most galaxies observed by \cite[van den Bosch et al. (2016))]{vandenBosch16} 
were of early type, with estimated nuclear masses $M_{\rm BH}$ $\ga$ 10$^8$\,M$_{\odot}$. Apparently, there is 
a rather narrow window for the occurrence of H$_2$O megamasers, which appear to be confined to nuclear regions 
with 10$^6$\,M$_{\odot}$ $<$ $M_{\rm BH}$ $<$ 10$^8$\,M$_{\odot}$. 

There are also recent 22\,GHz H$_2$O surveys with detections. Beside the RadioAstron measurements of known
sources with ultrahigh angular resolution, discussed by W.~A. Baan in this volume, a single-dish study to 
be mentioned in this context is that of \cite[Wagner (2013)]{Wagner13}. He targeted Seyfert 2 galaxies with high 
X-ray luminosity and high hydrogen column density, also including a few OH-absorbers. His detection rate, 
4 out of 37 galaxies, exceeds 10\% and therefore provides a good example on how to select promising sources. 
\cite[Yamauchi et al. (2017)]{Yamauchi17} chose galaxies with an absorbed 2\,keV continuum, a strong Fe 
6.4\,keV line and significant infrared emission. They detected three galaxies in a sample of 10 targets, 
an exceptionally high detection rate, with the caveat that their sample is comparatively small. Finally, 
\cite[Zhang et al. (2012)]{Zhang12} and \cite[Liu et al. (2017))]{Liu17} showed that almost exclusively 
Seyfert 2 galaxies with high radio continuum luminosities are exhibiting H$_2$O maser emission. This
led to a successful pilot study involving 18 sources (see \cite[Zhang et al. 2017)]{Zhang17} and the 
contribution by J.S. Zhang in this volume.

\section{The MCP}
At the core of the present extragalactic 22\,GHz H$_2$O maser research stands the Megamaser Cosmology Project 
(MCP) with the goals (1) to study accretion disk morphology, (2) to determine nuclear black hole masses, and 
(3) to constrain the Hubble constant to a precision of a few percent, avoiding any standard candles but using 
instead a direct geometric approach (see \cite[Herrstein et al. 1999]{Herrnstein99} for the first application 
of this technique). Introduced at the Stellenbosch meeting by \cite[Henkel et al. (2012)]{Henkel12}, three 
MCP articles had been published by that time, two on UGC~3789 and one on black hole masses (\cite[Reid et al. 
2009]{Reid09}; \cite[Braatz et al. 2010]{Braatz10}; \cite[Kuo et al. 2011)]{Kuo11}. In addition, \cite[Greene 
et al. (2010)]{Greene10} analyzed obtained black hole masses as a function of associated bulge masses. More 
recently, the number of MCP publications has tripled (\cite[Reid et al. 2013]{Reid13}, \cite[Kuo et al. 
2013, 2015]{Kuo13}, \cite[Pesce et al. 2015]{Pesce15}, \cite[Gao et al. 2016, 2017)]{Gao16i,Gao17}. These articles 
present detailed high resolution maps of a total of nine galaxies, most of them exhibiting the common 
disk-maser fingerprint. Detailed results are presented by J.~A. Braatz, in this volume. In addition, 
maser disk physics and upper magnetic field limits were discussed. Following \cite[Pesce et al. (2015)]{Pesce15}, 
the spiral shock model proposed by \cite[Maoz \& McKee (1998)]{MaozMcGee} is not consistent with the measured 
properties of the maser disks, since the so-called high velocity features, hundreds of km\,s$^{-1}$ off the 
parent galaxy's systemic velocity, do not show any of the predicted systematic velocity drifts. Limits obtained 
searching for Zeeman splitting typically reach 200-300\,mG (1$\sigma$), while the corresonding limit for 
NGC~1194 is 73\,mG.

\begin{figure}[b]
\includegraphics[width=13.0cm]{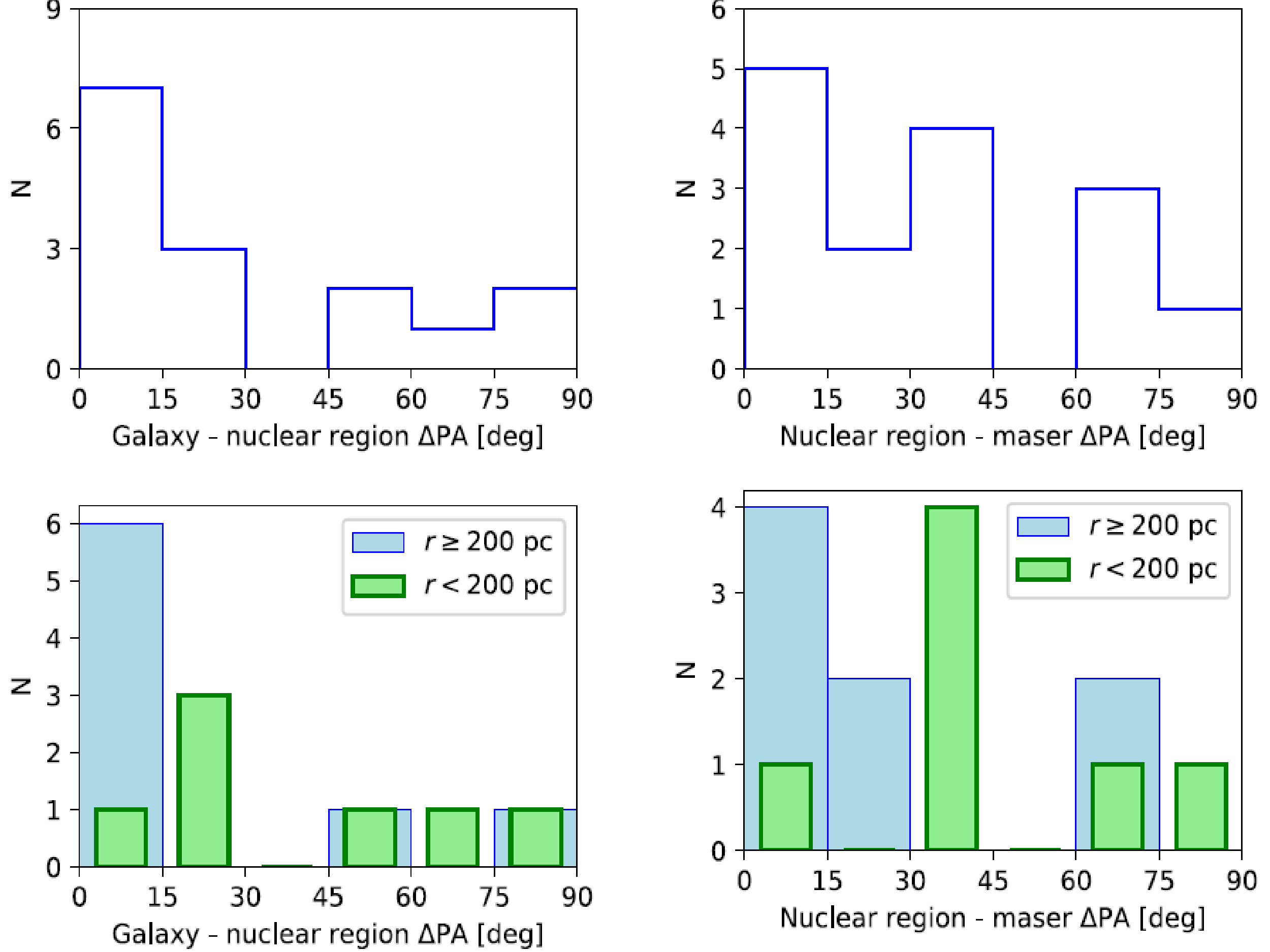}
\caption{Position angle (PA) differences between the angular momenta of the large scale disk versus the central 
region at radius $r$ $\approx$ 200\,pc (left) and the central region versus the pc-scale megamaser disk (right). 
The two lower panels also differentiate between nuclear regions $r$ $<$ 200\,pc and $>$200\,pc, still indicating
a possible alignment between the large scale disks and the $r$ $>$ 200\,pc regions, but not with the 
$r$ $<$ 200\,pc environment. Only the smaller one of two possible position angles is shown (the other one is 
180$^{\circ}$ minus the given angle).}
\label{pjanka}
\end{figure}

Beside disk morphology, $M_{\rm BH}$, and geometric distance, there are three major lines of research making 
use of the data supplied by the MCP: (1) A comparison of the resulting supermassive black hole masses with 
those derived by gas or stellar dynamics, bulge mass, total galactic mass, or mass within the central 
kiloparsec; (2) a comparison of position angle and inclination of the respective galaxies as a 
function of radius, starting from the outer large scale disk and proceeding to the smallest scales,
provided by the H$_2$O disk-masers, representing a nuclear disk viewed edge-on; (3) a test of the AGN 
paradigm, involving a supermassive nuclear engine, a jet, and, perpendicular to it, an accretion disk. 

(1) Comparisons of the mass of the nuclear engine with those of other galactic parameters reveal poor
correlations, thus leading to results which drastically differ from the tight correlation obtained 
for massive elliptical galaxies. Furthermore it appears (see Fig.~\ref{jenny}) that the masses of the 
supermassive nuclear engines of megamaser galaxies are below the expected correlation, while similarly 
sized spiral galaxies studied at optical or NIR wavelengths do not show this effect. \cite[Greene et al. 
(2016)]{Greene16} and \cite[L{\"a}sker et al. (2016))]{Laesker16} offer two most likely explanations
for this effect: Either the non-maser selected galaxies miss the low mass end of the BH distribution 
due to an inability to resolve their spheres of influence or the disk-maser galaxies preferentially 
occur in lower BH-mass environments. 

(2) The disk-maser galaxies with their known nuclear morphology allow for a unique comparison of position 
angles between largest and smallest scales. This reveals that the megamaser disks are neither aligned with 
the large scale disks of their parent galaxies nor with the morphology encountered at a galactocentric radius 
close to 100\,pc (\cite[Greene et al. 2010, 2013]{Green10}; \cite[Pjanka et al. 2016]{Pjanka16}). Fig.~\ref{pjanka} 
shows correlations between the large and $r$ $\approx$ 200\,pc scale (left panels) and between the $r$
$\approx$ 200\,pc and $\approx$ 1\,pc disk maser orientations. A differentiation between $r$ $>$ 200\,pc
and $<$ 200\,pc is also presented (lower panels).

\begin{figure}[b]
\includegraphics[width=13.0cm]{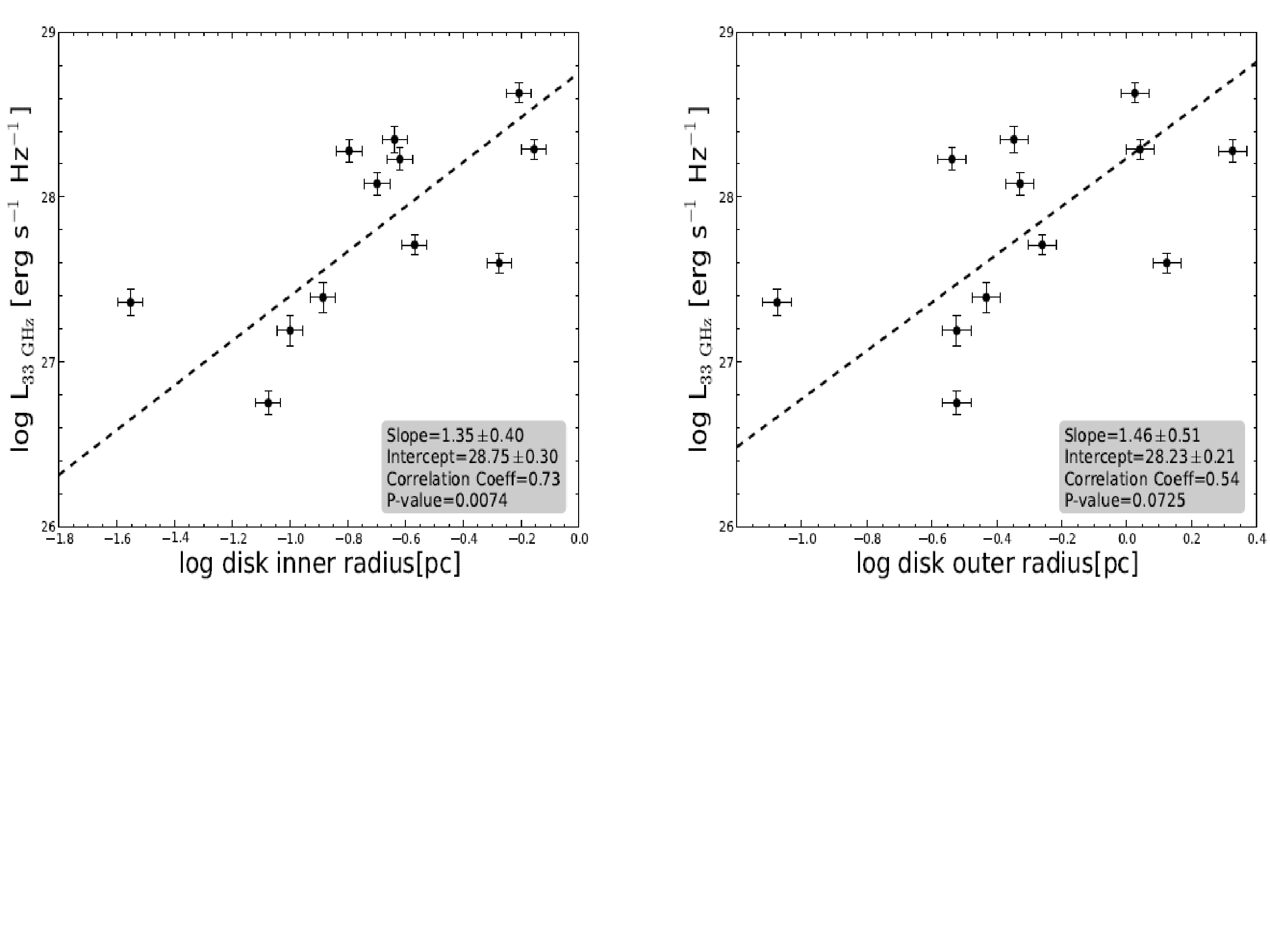}
\vspace{-3.4cm}
\caption{Maser disk inner (left) and outer (right) radius versus VLA 33\,GHz continuum luminosity. Slope,
intercept, correlation coefficient and p-values (likelihood that there is no correlation) are also 
given. From \cite[Kamali et al. (2017)]{Kamali17}.} 
\label{fateme1}
\end{figure}

(3) To become detectable, the 22\,GHz H$_2$O maser line requires gas fulfilling certain boundary 
conditions, like kinetic temperatures $T_{\rm kin}$ $\ga$ 300\,K and very high densities, $n$(H$_2$) 
$\ga$ 10$^7$\,cm$^{-3}$ (e.g., \cite[Kylafis \& Norman 1991]{Kylafis91}). Thus, the H$_2$O line is 
confined to a specific physical environment and is certainly not telling the entire story. To gain
deeper insights, radio continuum data are an excellent additional probe because they can be obtained 
with similar angular resolution and are also not affected by dust extinction. With the geometry being 
known, i.e. with nuclear disks viewed edge-on, radio continuum data are ideal to check the AGN paradigm: 
Are there jets and are these really two-sided, as it is expected for jets ejected parallel to the plane 
of the sky? Can we follow individual blobs, thus directly determining their speed? Can we detect emission 
from inside the maser disks? And are there correlations with the nuclear mass and/or the size of the 
maser disks? With this motivation dedicated continuum measurements at 33\,GHz (Very Large Array, VLA) 
and 5\,GHz (Very Long Baseline Array, VLBA) have been carried out. Some prelimiary VLBA results related 
to this ongoing project are presented by F. Kamali in this volume. Fig.~\ref{fateme1} shows the correlations
between the 33\,GHz VLA continuum flux densities and the inner and outer maser disk radii.

\section{A caveat when analyzing 22\,GHz H$_2$O maser emission}
\begin{figure}[b]
\begin{center}
\includegraphics[width=11.5cm]{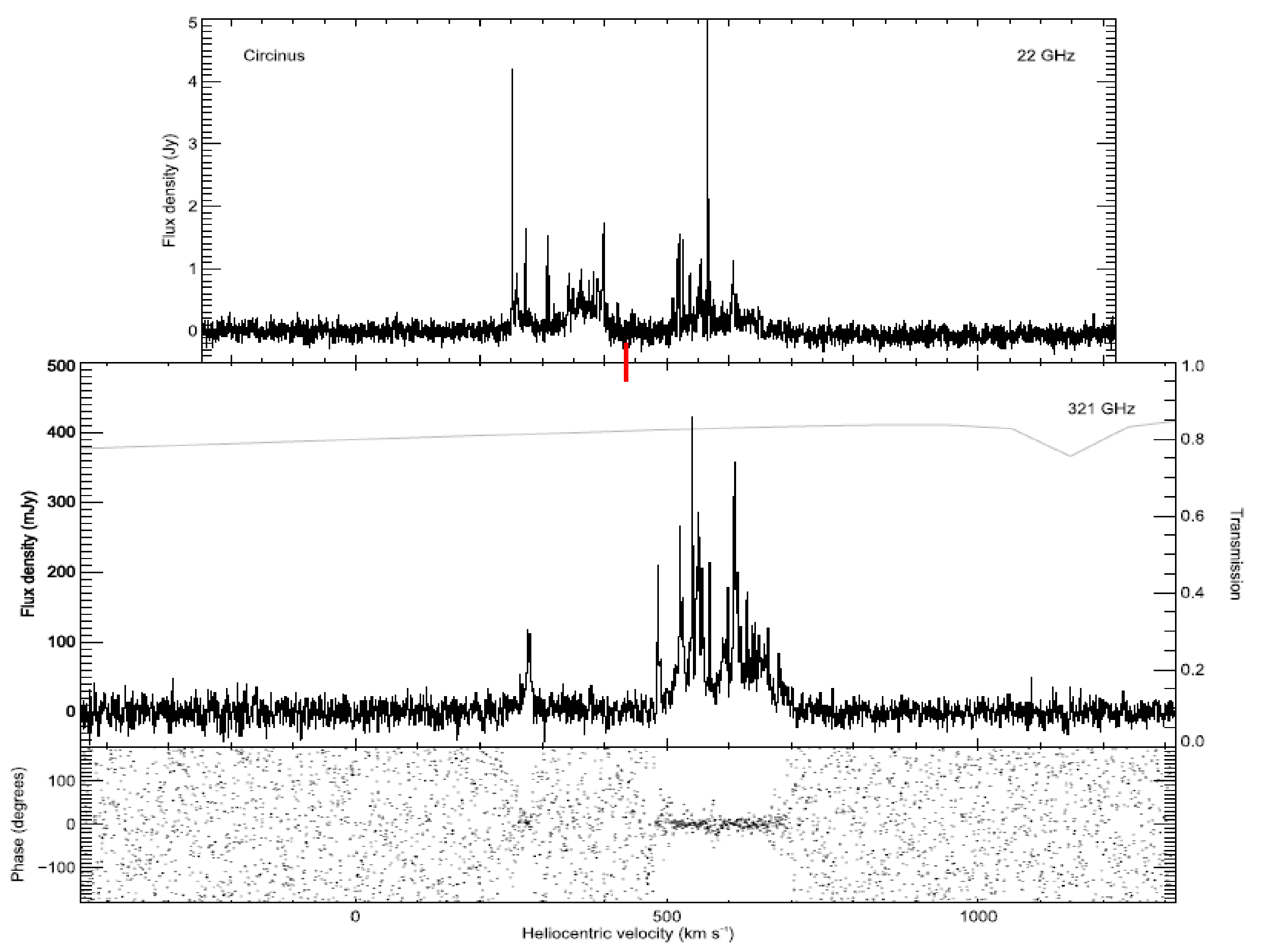}
\caption{22\,GHz (upper panel) and 321\,GHz (lower panel) H$_2$O maser spectra from the Circinus galaxy.
The bottom panel provides a phase plot for the calibrated 321\,GHz spectrum. The short vertical line 
connecting upper and lower panel denotes the systemic velocity. From \cite[Pesce et al. (2016)]{Pesce16}}. 
\label{pesce}
\end{center}
\end{figure}

Arp~220 with its extreme infrared luminosity ($L_{\rm IR}$ $\ga$ 10$^{12}$\,L$_\odot$) is often taken 
as the prototypical ULIRG. With the characteristic ratio of 10$^{-9}$ between (isotropic) 22\,GHz H$_2$O and 
infrared luminosity (e.g., fig.~9 in \cite[Henkel et al. 2005]{Henkel05}), we could thus expect strong 
maser emission with $L_{\rm H_2O}$ $\approx$ 1000\,L$_{\odot}$, yielding a flux density of $\approx$20\,mJy, 
if evenly distributed over a velocity range of 400\,km\,s$^{-1}$. Even in case of a deviation from this 
rule by a full factor of ten, 22\,GHz H$_2$O maser lines are characterized by narrow spikes (e.g., 
Fig.~\ref{pesce}), which would likely compensate for this effect. Furthermore, Arp~220 is characterized by 
a star formation rate amounting to a few 100\,M$_{\odot}$\,yr$^{-1}$ (e.g., \cite[Kennicutt 1998]{Kennicutt98}). 
Therefore there should be many star formation related H$_2$O masers, so many, that deviations from the 
10$^{-9}$-rule might be minimized.  In this respect it is noteworthy that 22\,GHz maser emission from this 
merging galaxy pair has only been detected recently (\cite[Zschaechner et al.  2016]{Zschaechner16}). The 
reason is absorption by the NH$_3$ ($J,K$) = (3,1) transition. The NH$_3$ (3,1) line at 22.234506\,GHz with 
levels 165\,K above the ground state and the H$_2$O line at 22.235080\,GHz, 645\,K above the ground state, 
are separated by only $\Delta V$ $\approx$ 8\,km\,s$^{-1}$. \cite[Zschaechner et al. (2016)]{Zschaechner16} 
find approximately --60\,Jy\,km\,s$^{-1}$ for the western and +60\,Jy\,km\,s$^{-1}$ for the eastern nucleus 
of Arp~220 (the absolute values are the same within $\approx$5\%), thus canceling but a negligible fraction 
of the total signal if viewed by a single-dish telescope. Toward the western nucleus NH$_3$ absorption dominates, 
while toward the eastern one emission, likely due to H$_2$O, is mainly seen. Apparently, when observing ULIRGs, 
high resolution data are mandatory, because in such galaxies gas densities and kinetic temperatures tend 
to be high, thus providing suitable conditions for the presence of non-metastable ammonia transitions 
interfereing with the 22\,GHz maser emission from H$_2$O if there is a continuum which can be absorbed. 
For an evaluation of the intensity of the NH$_3$ (3,1) line, then the detection of other non-metastable 
cm-wave NH$_3$ inversion transitions will be mandatory.

\section{Other H$_2$O maser lines}
With the 22\,GHz H$_2$O line detected in 178 galaxies (see the contribution by J. Braatz), there are also 
other promising lines to be observed. \cite[Humphreys et al. (2005)]{Humphreys05}  (NGC~3079) and 
\cite[Cernicharo et al. 2006]{Cernicharo06} (Arp~220) opened this field by observing the 183\,GHz 
3$_{13}$$\rightarrow$2$_{20}$ transition which is unlike the 6$_{16}$$\rightarrow$5$_{23}$ transition 
at 22\,GHz not connecting states 645\,K, but levels only 200\,K above the ground state. Thus emission 
is expected to be more widespread than at 22\,GHz, but is also affected by atmospheric obscuration, 
so that a non-zero redshift helps to enhance sensitivity. In recent years several additional studies have 
been carried out (\cite[Hagiwara et al. 2013b, 2016]{Hagiwara13b,Hagiwara16}; \cite[Galametz et al. 
2016]{Galametz16}, \cite[Humphreys et al. 2016]{Humphersy16}, \cite[Pesce et al. 2016]{Pesce16}),
focussing on this 183\,GHz line but also on the 321\,GHz 10$_{29}$$\rightarrow$9$_{36}$ transition, 
$\approx$1850\,K above the ground state and thus only tracing extremely highly excited gas. Studied sources
(see also the contribution by D. Pesce in this volume) are the Circinus galaxy, where the 321\,GHz line is 
covering a similar velocity range as the 22\,GHz line (perhaps even a slightly wider one, see Fig.~\ref{pesce}), 
and NGC~4945, where the 183\,GHz line is covering the entire velocity range, while the higher excitation gas 
sampled by the 321\,GHz transitions is only seen at the upper end of the galaxy's velocity range, near $V$ 
$\approx$ 700\,km\,s$^{-1}$. What is still missing here are high resolution observations of these lines 
allowing for a detailed comparison with the 22\,GHz data.

\section{1.7\,GHz OH versus 22\,GHz H$_2$O}
OH and H$_2$O megamasers appear to be mutually exclusive, likely because they trace very different physical environments.
\cite[Wiggins et al. (2016)]{Wiggins16} searched with the Green Bank Telescope (GBT) for luminous H$_2$O maser emission in OH 
megamaser galaxies and confirmed, after IC~694, with II\,Zw\,96 a second such object. Measuring instead OH in 
galaxies with strong H$_2$O masers using the Effelsberg telescope and the GBT, two of the sample galaxies were 
detected in OH, but in absorption. Details of this latter survey can be found in the contribution by E. Ladu, in this 
volume.

\begin{figure}[b]
\begin{center}
\includegraphics[width=13.5cm]{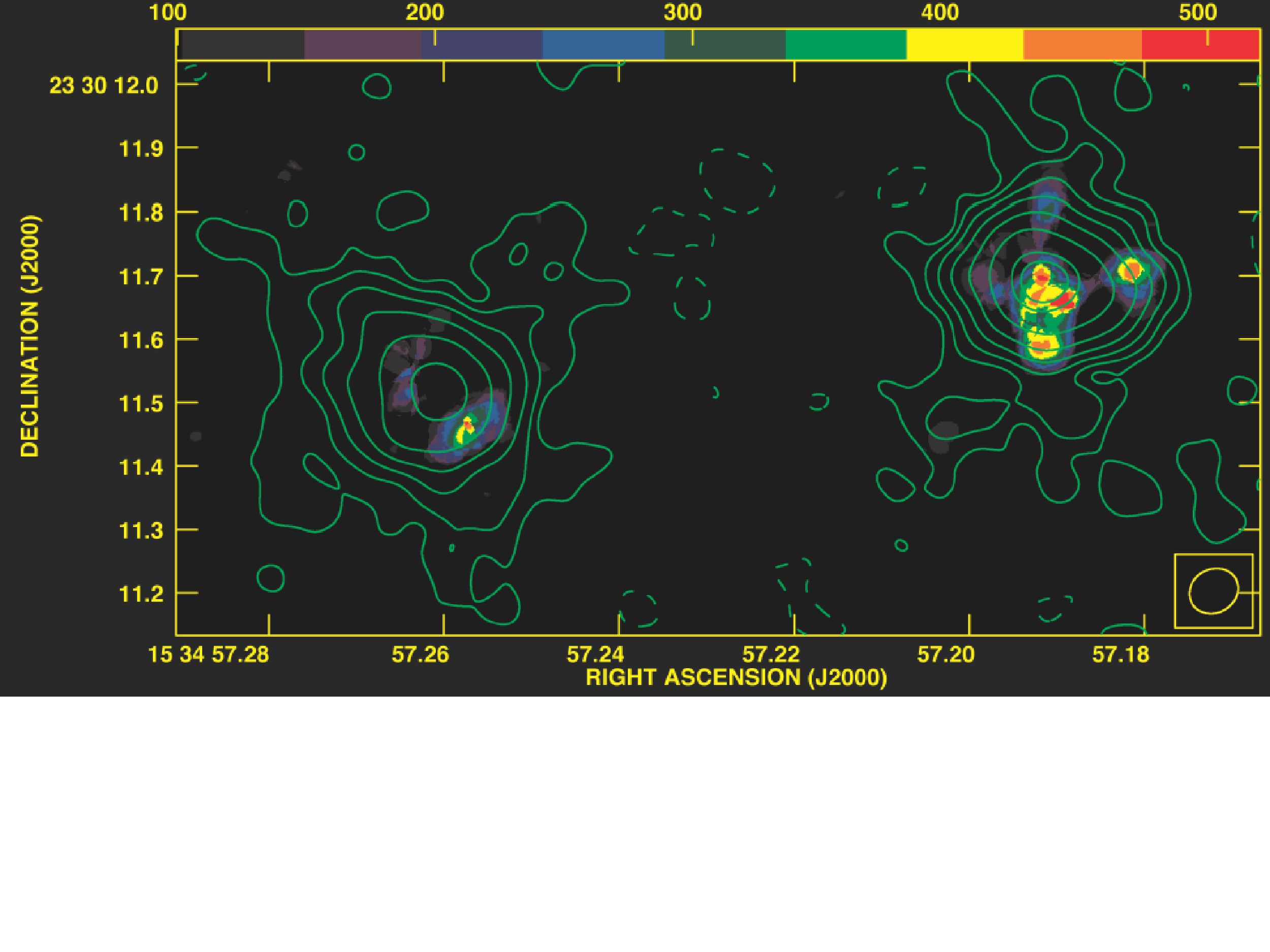}
\vspace{-3.0cm}
\caption{A composite map of the continuum (contours) and the 4.8\,GHz formadehyde emission of Arp~220 (from 
\cite[Baan et al. 2017]{Baan17}). Contour levels are 0.3 $\times$ (--1, 1, 2, 4, 8, 16, 32, 64, and 
80)\,mJy\,beam$^{-1}$ with peak flux densities of $\sim$13 and 30\,mJy\,beam$^{-1}$ for the eastern and western 
nucleus, respectively. A temperature scale in units of mJy\,\,km\,s$^{-1}$\,\,beam$^{-1}$ for the line emission is 
given at the upper edge of the image. The synthesized beam is indicated in the lower right.}
\label{willem}
\end{center}
\end{figure}

\section{Formaldehyde (H$_2$CO)}
\cite[Mangum et al. (2008, 2013)]{Mangum08} observed, with the Green Bank 100-m telescope, 56 star forming galaxies
in the 4.8 and 14.5\,GHz K-doublet transitions of formaldehyde (H$_2$CO). Both transitions were detected in 13
targets, mostly in absorption but sometimes also in emission. Applying Occam's Razor and assuming quasi-thermal and 
not maser radiation in case  of the emission lines, all data could be nicely fitted with a Large Velocity Gradient model, 
yielding densities in the range 10$^{4.5 ... 5.5}$\,cm$^{-3}$. Nevertheless, high resolution data from three galaxies 
with emission lines indicate that brightness temperatures can greatly exceed in some regions those expected in case of 
thermal emission, reaching values of 10$^{4...5}$\,K (\cite[Baan et al. 2017]{Baan17}). The three galaxies are IC~860, 
IR\,15107+0724 and Arp~220 (see Fig.~\ref{willem} for an image). As a consequence, it would be interesting to 
evaluate in how far this affects the analysis by \cite[Mangum et al. (2080, 2013)]{mangum08}, which is mainly but 
not entirely based on H$_2$CO 4.8 and 14.5\,GHz absorption lines.

\section{Methanol, HC$_3$N, and HCN}
In the Galaxy, the so-called Class II 6.7\,GHz transition exhibits the strongest methanol maser lines, being
directly associated with sites of massive star formation (e.g., \cite[Menten 1991]{Menten91}). Searches for 
this line in extragalactic space were, however, not very successful, only providing maser detections in the 
Magellanic Clouds and the Andromeda galaxy (see Sect.\,2). After unsuccessfully searching for Class II lines
much stronger than those encountered in the Galaxy, Class I masers, being clearly less conspicuous in the Galaxy
and less directly associated with sites of massive star formation, have become part of a recent survey. 
And here, surprisingly, emission much more luminous than the corresponding Galactic masers could be found
(\cite[Ellingsen et al. 2014, 2017a]{Ellingsen14}; \cite[Chen et al. 2015, 2016]{Chen15}; \cite[McCarthy et 
al. 2017]{McCarthy17}). What makes these detections peculiar is that the masers are not associated with 
the very center of their parent galaxies, but are instead detected in the outskirts of their nuclear environments.
A detailed account of these new findings, including the possible detection of an HC$_3$N $J$ = 4$\rightarrow$3 
maser (\cite[Ellingsen et al. 2017b]{Ellingsen17b}), is given by the contributions of S. Ellingsen, X. Chen, and 
T. McCarthy in this volume. The presence of a weakly masing  HCN $J$ = 1$\rightarrow$0 line right at the center of 
the whirlpool galaxy M~51, in a region with relatively weak CO emission, has been proposed by Matsushita et al. 
(2015).

\section{The bright future}
Interestingly, and possibly for the first time in this series of meetings, extragalactic OH was not a topic of 
much discussion. However, this will likely change in the forthcoming years. With Apertif (Aperture Tile in Focus), 
MeerKat (Meer Karoo Array Telescope), ASKAP (Australia Square Kilometer Array Pathfinder) and FAST (Five 
hundred meter Aperture Spherical Telescope), new OH surveys sometimes piggybacking on H{\sc i} measurements, 
will cover significant parts of the sky. This will greatly help in obtaining new detections of luminous OH 
masers and to reach eventually an independent estimate of the number of merger galaxies as a function of redshift. 

With the Very Large Array (VLA) and possibly also with the Atacama Large Millimeter/submillimeter Array (ALMA), 
new extragalactic Class I methanol maser surveys can be carried out. Here we are still near the start. While 
the standard lines near 36 and 44\,GHz have been measured in a small number of galaxies, higher frequency 
Class I masers detectable with ALMA have, to our knowledge, not even been touched.

With respect to 22\,GHz H$_2$O masers, we note that the MCP is close to completion. We can expect a final
Hubble constant deduced from this survey with an uncertainty of only a few percent during the next one 
or two years. However, this will be by no means the end of H$_2$O megamaser research. Not only other H$_2$O 
lines are and will remain attractive. The 22\,GHz maser line itself may remain at the center of interest. 
The reason is that detection rates of future 22\,GHz maser surveys have the potential to go up dramatically.
So far, average detection rates are well below 10\%. Detection rates of disk-masers are even close to 
or less than 1\%. All this may change with the introduction of new criteria, either led by X-ray spectroscopy, 
by the radio continuum luminosity, or by a combination of the two methods. At the same time, the James Webb Space
Telescope (JWST), ALMA and NOEMA (NOrthern Extended Millimeter Array) may provide new constraints to 
the masses of nuclear engines, to be compared with those deduced from the maser-disk data.

\end{document}